\documentclass[12pt,a4paper]{article}
\usepackage{jcappub}

\pdfoutput=1 

\DeclareMathOperator{\arsinh}{arsinh}
\DeclareMathOperator{\Tr}{Tr}
\DeclareMathOperator{\Det}{Det}

\usepackage[compat=1.1.0]{tikz-feynman}

\usetikzlibrary{decorations.markings,arrows.meta}

\tikzfeynmanset{ with arrow/.style = {
   decoration={
     markings,
     mark=at position .55
          with {\arrow{Stealth[scale=1.4]}}
     },
   postaction=decorate}
}

\title{Interactions of the scalaron dark matter in $f (R)$ gravity}

\author[a]{Yuri Shtanov}
\author[b]{and Yurii Sheiko} 

\affiliation[a]{Bogolyubov Institute for Theoretical Physics, \\ Metrologichna St.\@ 14-b, Kiev 03143, Ukraine} %
\affiliation[b]{Department of Physics, Taras Shevchenko National University of Kyiv, \\ Academician Glushkov Ave.\@ 3, Kiev 03022, Ukraine} %

\emailAdd{shtanov@bitp.kyiv.ua}
\emailAdd{ura.sheiko@gmail.com}

\abstract{In $f(R)$ gravity, the scalaron\,---\,a scalar degree of freedom arising from modification of General Relativity\,---\,could account for all dark matter in the universe if its mass lies in the meV--MeV range. In this work, we revisit the scalaron's interactions with Standard Model particles, assuming their minimal coupling to gravity. In particular, we provide a detailed calculation of the scalaron's decay rate into two photons\,---\,a one-loop process of significant interest that has been the subject of discrepancies in the literature. We demonstrate that a direct evaluation of loop diagrams with appropriate regularisation eliminates the ambiguities inherent in methods relying on Jacobians from field redefinitions. Assuming the scalaron constitutes all of dark matter, we calculate the average cosmological background radiation produced by its decays into photons. We also estimate the contribution of primordial scalarons emitted in the hot early universe to the present dark matter density and find it to be negligible. Our results support all key aspects of the original scenario, in which scalaron dark matter behaves as a coherently oscillating field.}


\arxivnumber{2505.00324}

\begin{document} 
\maketitle
\flushbottom

\section{Introduction}

One of the dark matter candidates is the scalar degree of freedom (the scalaron) of $f (R)$ gravity \cite{Capozziello:2006uv, Nojiri:2008nt, Cembranos:2008gj, Cembranos:2015svp, Corda:2011aa, Katsuragawa:2016yir, Katsuragawa:2017wge, Yadav:2018llv, Parbin:2020bpp, Shtanov:2021uif, KumarSharma:2022qdf} (see also reviews \cite{Sotiriou:2008rp, DeFelice:2010aj, Nojiri:2010wj}). Among various proposals in this framework, we focus here on a straightforward suggestion, pioneered in \cite{Cembranos:2008gj}, in which a key role is played by the term quadratic in curvature. In this model, dark matter consists of a classical scalaron field oscillating near the minimum of its potential. The oscillating scalaron field clusters like cold dark matter, contributing to the formation of large-scale structure with its typical gravitational effects. A notable advantage of this theory is its simplicity: it introduces no new fundamental fields, aside from the fact that the metric acquires an additional degree of freedom. Assuming the standard (minimal) coupling of the metric to the matter fields, the model has only one essential free parameter\,---\,the mass of the scalaron\,---\,making it highly predictive. The main observational implications of the model have been discussed in \cite{Cembranos:2008gj, Cembranos:2015svp}, and are reviewed and updated in this paper. Thus, due to the weak interaction with matter, the scalaron is able to decay into photons or electron-positron pairs, which allows one to obtain an upper limit on the scalaron mass.

While various aspects of the scalaron's interactions with matter have been studied in the literature, certain features of its coupling to Standard Model fields warrant a more detailed examination. This work focuses on the scalaron's decay into two photons\,---\,a one-loop process of particular interest that has been subject to some controversy in the literature. We provide an explicit derivation of the decay rate and show that applying an appropriate regularisation to the relevant loop diagrams resolves the ambiguities arising in calculations that rely on regularised Jacobians from spinor field redefinitions. Assuming the scalaron constitutes all of dark matter, we calculate the average cosmological background radiation produced by its decays into photons.

The issue of initial conditions for the scalaron was thoroughly addressed in our recent papers \cite{Shtanov:2022xew, Shtanov:2024nmf}. However, one important aspect that has not been explored is the fraction of thermal scalarons emitted by the hot cosmic plasma in the early universe. In this work, we estimate their contribution to the total scalaron dark matter density and show that it is, in fact, negligible. This supports the scenario in which scalaron dark matter behaves as a coherently oscillating classical field, as originally proposed in \cite{Cembranos:2008gj}.

The paper is organised as follows. In section~\ref{sec:scalaron}, we introduce the $f(R)$ gravity model and outline the role of the scalaron as a dark matter candidate. Section~\ref{sec:coup} describes the scalaron's coupling to Standard Model fields. In section~\ref{sec:decay}, we compute the average cosmological background radiation resulting from scalaron decays into photons. In section~\ref{sec:thermal}, we estimate the contribution of primordial thermal scalarons to the present dark matter density. Our findings are summarised in section~\ref{sec:summary}. Appendix~\ref{app:decay} provides the expression for the one-loop effective interaction between the scalaron and the electromagnetic field, while appendix~\ref{app:redef} addresses spinor field redefinitions and the associated ambiguities.

\section{Scalaron of $f(R)$ gravity as dark matter}
\label{sec:scalaron}

The $f (R)$ gravity theory assumes that the Lagrangian for gravity can be expanded in the form of power series in the scalar curvature $R$:
\begin{equation} \label{Sgs}
L_g = \frac{M^2}{3} f (R) \, , \qquad f (R) = - 2 \Lambda + R + \frac{R^2}{6 m^2} + \ldots \, .
\end{equation} 
Here, $M = \sqrt{3 / 16 \pi G} \approx 2.98 \times 10^{18}\, \text{GeV}$ is a conveniently normalised Planck mass, and $\Lambda \approx \left( 3 \times 10^{-33}\,\text{eV} \right)^2$ is the cosmological constant in the natural units $\hbar = c = 1$. We are working in the metric signature $(-, +, +, +)$.

The theory involving terms up to $R^2$ in \eqref{Sgs} corresponds to the Starobinsky inflationary model \cite{Starobinsky:1980te, Vilenkin:1985md}, where setting $m \simeq 10^{-5} M$ yields a primordial power spectrum that is consistent with current observations \cite{Planck:2018jri}. In this context, the quadratic curvature term can be interpreted as a quantum correction to the effective action for gravity, emerging from the integration of certain matter degrees of freedom. When applying this model to describe dark matter, instead of inflation, the mass $m$ must be constrained to lie between the meV and MeV range, as will be discussed below. The mass $m$ should then be regarded as a genuine constant in the action for gravity. The assumed smallness of $m$ (or, in other words, the large dimensionless factor $M^2 / 18 m^2 \sim 10^{41}$--$10^{58}$ of the $R^2$ term) can be viewed on the same footing as the extreme relative smallness of the cosmological constant $\Lambda$ in the gravitational action, the true reason for which is also unknown. 

To identify the new scalar degree of freedom, one can proceed from the Jordan frame to the Einstein frame. We first write the action with Lagrangian \eqref{Sgs} in the form
\begin{equation}\label{Sg1}
S_g = \frac{M^2}{3} \int d^4 x \sqrt{-g}\, \bigl[ \Omega R - U (\Omega) \bigr] \, ,
\end{equation}
where $\Omega$ is a new dimensionless scalar field, and the function $U (\Omega)$ is chosen so that variation with respect to $\Omega$ and its substitution into the action returns the original action: 
\begin{align} \label{inv1}
U' (\Omega) &= R \quad \Rightarrow \quad \Omega = \Omega (R) \, , \\
f (R) &= \bigl[ \Omega R - U (\Omega) \bigr]_{\Omega = \Omega (R)} \, . \label{inv2}
\end{align}
Thus, $f (R)$ is the Legendre transform of $U (\Omega)$, and vice versa. 

As a next step, one performs a conformal transformation\footnote{In quantum field theory, this transformation is typically referred to as the Weyl transformation. We use the terminology commonly adopted in the context of gravity theory (see, e.g., \cite{Wald:1984}).} in \eqref{Sg1}:
\begin{equation} \label{om}
g_{\mu\nu} = \Omega^{-1}\, \widetilde g_{\mu\nu} \, , \qquad \Omega = e^{\phi / M} \, ,
\end{equation}
where $\phi$ is a new field (the scalaron) parametrising $\Omega$, and we use a tilde to denote all metric-related quantities in the new (Einstein) frame. Action \eqref{Sg1} then becomes that of the Einstein gravity with a scalar field (scalaron) $\phi$. The Lagrangian in the Einstein frame is 
\begin{equation}\label{Sg3}
L_g =  \frac{M^2}{3} \widetilde R - \frac12 \widetilde g^{\mu\nu} \partial_\mu \phi \partial_\nu \phi - V (\phi) \, ,
\end{equation}
where the scalaron potential $V (\phi)$ is calculated from \eqref{Sg1}:
\begin{equation} \label{V}
V (\phi) = \frac{M^2}{3} e^{- 2 \phi / M} U \bigl( e^{\phi / M} \bigr) \, .
\end{equation}

It is easy to establish that the scalaron potential has extrema, with $V' (\phi) = 0$, at the Jordan-frame values of $R$ that satisfy
\begin{equation}
R f' (R) = 2 f (R) \, .
\end{equation}
The scalaron mass squared, $m_\phi^2 = V'' (\phi)$, at such an extremum is given by
\begin{equation}
m_\phi^2 = \frac13 \left[ \frac{1}{f''(R)} - \frac{R}{f'(R)} \right] = \frac13 \left[ \frac{1}{f''(R)} - \frac{R^2}{2 f (R)} \right] \, .
\end{equation}
If $m_\phi^2 > 0$, then this is a local minimum. As can be seen from \eqref{V}, the scalaron potential typically varies on the scale of the Planck mass $M$.  Hence, in the neighbourhood of the minimum, it is well approximated by a quadratic form on field scales much smaller than $M$.  

For a small cosmological constant, $\Lambda \ll m^2$, the theory has a local minimum at $\phi / M \approx 4 \Lambda / 3 m^2$, corresponding to $R \approx 4 \Lambda$ in the Jordan frame, and the scalaron mass at this minimum is $m_\phi^2 = m^2 + {\cal O} (\Lambda)$. In what follows, we neglect the small cosmological constant in \eqref{Sgs}, which is responsible for dark energy but not for dark matter. In this approximation, the local minimum is situated at $\phi = 0$ (corresponding to $R = 0$ in the Jordan frame), and the scalaron mass at this minimum is $m_\phi = m$. 

The minimal non-trivial model \cite{Starobinsky:1980te, Vilenkin:1985md} is described by
\begin{equation}\label{fstar}
f (R) = R + \frac{R^2}{6 m^2} \, .
\end{equation}
In the Einstein frame, it produces Lagrangian \eqref{Sg3} with
\begin{equation} \label{Vstar}
V (\phi) = \frac12  m^2 M^2 \left( 1 - e^{- \phi / M} \right)^2 \, . 
\end{equation}
This potential has an infinitely extended plateau at $\phi \gg M$. 

\begin{figure}
\begin{center}
\includegraphics[width=.7\textwidth]{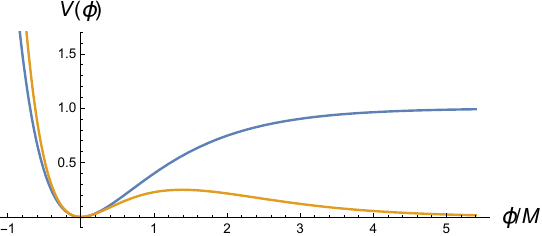}
\caption{Scalaron potentials \eqref{Vstar} (blue) and \eqref{Vrat} (orange) are plotted in units $M^2 m^2/2$. In the region $|\phi|/M \ll 1$, they both are approximated by the quadratic form with mass $m$. \label{fig:pot}}
\end{center}
\end{figure}

Scalaron potentials corresponding to $f (R)$ with higher powers of $R$ in their expansion typically have different behaviour at large values of $\phi > 0$, exhibiting a `hilltop' or `tabletop' shape \cite{Shtanov:2022pdx}.  As an example, consider
\begin{equation} 
f (R) = \frac{R}{1 - R / 6 m^2} = R + \frac{R^2}{6 m^2} + \frac{R^3}{36 m^4} + \ldots \, .
\end{equation}
The branch of the theory containing the stable critical point $R = 0$ has scalaron potential
\begin{equation} \label{Vrat}
V (\phi) = 2 M^2 m^2 e^{- \phi / M}  \left( 1 - e^{- \phi / 2 M} \right)^2 \, .
\end{equation}
It has a local maximum at $e^{\phi/M} = 4$ and exponentially decreases as $\phi \to \infty$. Potentials \eqref{Vstar} and \eqref{Vrat} are plotted in figure~\ref{fig:pot}. 

A common feature of all such potentials in a theory containing the quadratic term $R^2 / 6 m^2$ is that they are well approximated by the quadratic form with mass $m$ at the origin. The minimal coupling of the metric to matter in the Jordan frame results in a universal coupling of the scalaron to matter fields, collectively denoted as $\Psi$, with the Lagrangian density
\begin{equation}
{\cal L}_\text{m} \left( e^{- \phi / M} \widetilde g_{\mu\nu}, \Psi \right) \, .
\end{equation}
Here, the fields $\Psi$ are those of the Jordan frame, while the metric is transformed to the Einstein frame. Note that it is the Jordan-frame metric $g_{\mu\nu} = e^{- \phi / M} \widetilde g_{\mu\nu}$ that plays the role of the `observable' metric in the Einstein frame \cite{Dicke:1961gz, Faraoni:2006fx, Shtanov:2022wpr}, and the presence of the scalaron $\phi$ in this metric then produces an additional gravitational force of Yukawa type, with the total gravitational potential per unit gravitating mass \cite{Stelle:1977ry}
\begin{equation}
\Phi_\text{grav} = - \frac{2 G}{r} \left( 1 + \frac13 e^{- m r} \right) \, .
\end{equation}
Non-observation \cite{Kapner:2006si, Adelberger:2006dh} of such additional Yukawa forces between non-relativistic masses at small distances leads to a lower bound on the scalaron mass (see also \cite{Cembranos:2008gj, Cembranos:2015svp, Perivolaropoulos:2019vkb})
\begin{equation}\label{mlow}
m \geq 2.7~\text{meV} \quad \text{at 95\% C.L.}
\end{equation}

If the scalaron forms all of dark matter, then the value of this field in the early universe should be very close to the minimum of its potential.  Indeed, for the amplitude $\phi_\text{a}$ of the scalaron oscillations, one has the estimate
\begin{equation}\label{ampl}
\frac{\phi_\text{a}}{M_\text{p}} = \frac{\sqrt{2 \rho_\phi}}{m M_\text{p}} = \frac{2 \sqrt{\Omega_\text{\tiny DM}} H_0}{m} \sqrt{\frac{\rho_\phi}{\overline \rho_\phi}}\, \simeq \, 10^{- 30} \sqrt{\frac{\rho_\phi}{\overline \rho_\phi}}\, \frac{\text{meV}}{m}\, \lll \, 1 \, .
\end{equation}
Here, $\rho_\phi = m^2 \phi_a^2 / 2$ is the local energy density of the scalaron dark matter, and $\overline \rho_\phi$ is its cosmological average today. The quantities $\Omega_\text{\tiny DM}$ and $H_0$ are the current cosmological parameter for dark matter and the Hubble constant, respectively. 

In the very early universe, the scalaron field is initially frozen due to Hubble friction, beginning to oscillate once the Hubble friction parameter $3H$ decreases to the value of its mass $m$. For the early hot universe, this condition reads
\begin{equation}\label{Hm}
3 H_\text{i} \equiv \sqrt{\frac{3 N_\text{i}}{5}} \frac{\pi T_\text{i}^2}{M} \simeq m \, ,
\end{equation}
where $N_\text{i} \approx 100$ is the number of relativistic degrees of freedom in thermal equilibrium at this moment. Thus, $T_\text{i} \propto m^{1/2}$, and $\left(\phi_\text{a} \right)_\text{i} / M \propto m^{-1/4}$.  Given the lower bound \eqref{mlow} on $m$, the estimate \eqref{ampl} suggests that $\left(\phi_\text{a} \right)_\text{i} / M \lesssim 4 \times 10^{-7}$. Therefore, the scalaron is expected to remain very close to zero throughout the entire cosmological history.

\section{Coupling to the Standard Model}
\label{sec:coup}

We assume that matter is described by the Standard Model minimally coupled to the metric in the Jordan frame. Note that most of the Standard Model action is classically conformally invariant (with proper conformal transformation of the matter fields). The only part that breaks classical conformal invariance is the Higgs sector and the neutrino sector extending the Standard Model if it contains Majorana mass terms. 

The Higgs sector has the Lagrangian 
\begin{equation}\label{Sh}
L_\text{\tiny H} = - g^{\mu\nu} \left( D_\mu \Phi \right)^\dagger D_\nu \Phi - \frac{\lambda}{4} \left( 2 \Phi^\dagger \Phi - v^2 \right)^2 \, .
\end{equation}
Here, $D_\mu$ is the gauge covariant derivative involving the U(1)$_Y$ and SU(2)$_L$ electroweak gauge fields and acting on the Higgs doublet $\Phi$, and $v \approx 246\, \text{GeV}$ is the symmetry-breaking parameter. The Higgs boson has mass $m_\text{\tiny H} = \sqrt{2 \lambda} v \approx 125\,\text{GeV}$ in the Standard Model, so that $\lambda \approx 0.13$.

After the conformal transformation \eqref{om}, the Lagrangian becomes (we remember the factor $\sqrt{-g}$ in the Lagrangian density)
\begin{equation}\label{Shn}
L_\text{\tiny H} = - e^{- \phi / M} \widetilde g^{\mu\nu} \left( D_\mu \Phi \right)^\dagger D_\nu \Phi - \frac{\lambda}{4} e^{- 2 \phi / M} \left( 2 \Phi^\dagger \Phi - v^2 \right)^2 \, .
\end{equation}
We observe the appearance of non-renormalisable interactions of the scalaron $\phi$ with the Higgs field, which are suppressed by inverse powers of the large Planck mass $M$. 

Similarly, in the new conformal frame, the scalaron has direct couplings to both the kinetic and mass terms of fermionic fields. This complication can be circumvented by noting that the spinor kinetic terms in the action are conformally invariant, provided the spinors are appropriately rescaled. Hence, in the rest of the Standard Model, one can also perform the conformal transformation of the Dirac spinor fields, 
\begin{equation}\label{psicon}
\psi = \Omega^{3/4} \widetilde \psi = e^{3 \phi / 4 M} \widetilde \psi \, ,
\end{equation}
which brings their kinetic terms to the canonical form in the Einstein frame. This transformation belongs to the class of point transformations that preserve all symmetries. In appendix~\ref{app:redef}, we show that this transformation satisfies the field-redefinition equivalence theorems \cite{Chisholm:1961tha, Kamefuchi:1961sb, Divakaran:1963yxz, Arzt:1993gz, Cohen:2024fak}\,---\,at least at the one-loop level\,---\,thereby leaving the $S$-matrix invariant. 

Under transformation \eqref{psicon}, a typical term with the Yukawa coupling to the Higgs field transforms as 
\begin{equation}\label{Yukawa}
\overline \Psi_L \Phi \psi_R\, \sqrt{- g} \, \to \, e^{- \phi / 2 M}\, \overline{\widetilde \Psi}_L \Phi \widetilde \psi_R\, \sqrt{- \widetilde g} \, .
\end{equation}
Here, $\Psi_L$ is a left-handed doublet and $\psi_R$ is a right-handed singlet. Possible Majorana mass terms for neutrinos will transform in a similar way:
\begin{equation}\label{Majorana}
M_\psi \overline{\psi^c_R} \psi^{}_R\, \sqrt{- g} \, \to \, e^{- \phi / 2 M} M_\psi \overline{\widetilde \psi^c_R} \widetilde \psi^{}_R\, \sqrt{- \widetilde g} \, .
\end{equation}

In the unitary gauge for the Higgs field, we will have 
\begin{equation}\label{higgs}
\Phi = \frac{1}{\sqrt{2}} \begin{pmatrix} 0 \\ h \end{pmatrix} = \frac{1}{\sqrt{2}} \begin{pmatrix} 0 \\ v + \chi \end{pmatrix}\, ,
\end{equation}
where $\chi$ is the deviation of the Higgs field $h$ from its vacuum expectation value $v$. Then the standard Yukawa interaction $- \gamma h \bar \psi \psi$ between the Higgs field $h$ and a fermion $\psi$, where $\gamma$ is the Yukawa coupling constant, will\,---\,according to \eqref{Yukawa} and to first order in $\phi$\,---\,induce the following coupling:
\begin{equation}\label{psicoup}
\frac{\gamma h}{2 M} \phi\, \overline{\widetilde \psi} \widetilde \psi = \frac{m_\psi (h)}{2 M} \phi\, \overline{\widetilde \psi} \widetilde \psi \, ,
\end{equation}
where $m_\psi (h) = \gamma h$ is the temperature-dependent fermion mass. Similar coupling is observed in the Majorana mass terms \eqref{Majorana}. Likewise, the coupling between the Higgs field and the vector gauge fields in \eqref{Shn}, to first order in $\phi$, will generate the couplings\footnote{The tensor indices here and below are contracted by the Einstein metric $\widetilde g^{\mu\nu}$.}
\begin{equation}\label{gcoup}
\frac{m_W^2 (h)}{M} \phi\, W_\mu^+ W^{-\mu} + \frac{m_Z^2 (h)}{2 M} \phi\, Z_\mu Z^\mu
\end{equation} 
of the scalaron to the vector bosons $W^\pm$ and $Z^0$. These couplings would allow for the scalaron decays into other particles provided its mass $m$ is sufficiently large. 

In particular, interaction \eqref{psicoup} allows for the scalaron decays into electron-positron pairs if $m > 2 m_e$. For the scalaron, with interaction \eqref{psicoup}, we have
\begin{equation}\label{Gammaep}
\Gamma_{\phi\, \to\, e^+ e^-} = \frac{m_e^3}{16 \pi M^2} \frac{\left( r_e^2 - 1 \right)^{3/2}}{r_e^2} = 4.54 \times 10^{-25}\, \frac{\left( r_e^2 - 1 \right)^{3/2}}{r_e^2}~\text{s}^{-1} \, , 
\end{equation}
where $r_e = m / 2 m_e \geq 1$. 

For the scalaron representing all of dark matter in the universe, an upper bound on the scalaron mass can be obtained from the observed 511~keV emission line from the Galactic Centre \cite{Cembranos:2008gj, Cembranos:2015svp}, which is consistent with a spectrum from electron-positron annihilation. A possible source of positrons in this case could be a decaying dark matter with mass $1~\text{MeV} \lesssim m_\text{\tiny DM} \lesssim 10~\text{MeV}$ satisfying the relation (see \cite{Cembranos:2008gj, Cembranos:2015svp} and references therein)
\begin{equation}\label{mdm}
m_\text{\tiny DM} \, \simeq \, \Omega_\text{\tiny DM} h_{100}^2\, \Gamma_{\text{\tiny DM} \to e^+ e^-} \cdot \left( \text{0.2--4} \right) \times 10^{27}~\text{s MeV} \, ,
\end{equation}
where $h_{100} = H_0 / 100$~{km/s~Mpc}, and $\Gamma_{\text{\tiny DM} \to e^+ e^-}$ is the decay rate of a dark-matter particle into an electron-positron pair. The uncertainty in \eqref{mdm} comes from modelling the dark matter halo profile.  For the Planck-2018 value of $\Omega_\text{\tiny DM} h_{100}^2 = 0.120 \pm 0.001$ \cite{Planck:2018vyg}, equations \eqref{Gammaep} and \eqref{mdm} then give
\begin{equation}\label{mup}
1.04~\text{MeV} \, \lesssim \, m \lesssim 1.15~\text{MeV} \, ,
\end{equation}
with the right inequality being roughly the upper bound on the scalaron mass if it constitutes all of dark matter.

\section{Photons from the scalaron decays}
\label{sec:decay}

Interactions of the form \eqref{psicoup} and \eqref{gcoup} enable scalaron decays into photons via quantum loops. These decays occur in perturbative regime, without the effects of induced radiation (or, equivalently, development of parametric resonance) \cite{Shtanov:2024nmf}. The corresponding decay rate and lifetime, in the neighbourhood of $m = 2 m_e$, are then calculated to be (see appendix~\ref{app:decay})
\begin{equation} \label{Gamma1}
\Gamma_{\phi\, \to\, \gamma\gamma} \approx 5.2 \times 10^{- 30} \left( \frac{m}{\text{MeV}} \right)^3\, \text{s}^{-1} \approx \left[ 1.9 \times 10^{29} \left( \frac{\text{MeV}}{m} \right)^3\, \text{s} \, \right]^{-1} \, ,
\end{equation}
which is close to the value quoted in \cite{Cembranos:2008gj}. This can be compared with the age of the universe, $1.4 \times 10^{10}~\text{yr} \approx 4.2 \times 10^{17}~\text{s}$. 

In this section, we estimate the spatially averaged spectrum of cosmological photons originating from this source, assuming a spatially homogeneous universe. By $f (p) d p$ we denote the comoving number density of photons with comoving absolute momentum $p$ in the interval $d p$. After photon production, this quantity remains constant in a homogeneous universe, provided photon spectral diffusion and absorption can be neglected. To demonstrate that these processes can indeed be ignored, we estimate the mean free path $\ell_c \simeq {1}/{\sigma_c n_e}$ for Compton scattering off free electrons with average number density $n_e$, where $\sigma_c \simeq \pi \alpha^2 / m_e^2$ is the scattering cross-section. The average number density of electrons in the universe is estimated to be
\begin{equation}
n_e \approx 2 \times 10^{-7}\, (1 + z)^3~\text{cm}^{-3} \, ,
\end{equation}
where $z$ is the cosmological redshift. Hence, we obtain
\begin{equation}
\ell_c \simeq 2 \times 10^{31}\, (1 + z)^{-3}~\text{cm} \, .
\end{equation}
This becomes comparable to the Hubble radius $H^{-1} (z)$ at $z \approx 80$, well beyond the reionisation epoch, when the universe is transparent due to its neutrality. Therefore, we can safely neglect the effects of Compton scattering in the late-time universe.

Now we can calculate the distribution function $f (p)$. During the photon production, the photon comoving momentum $p = a m/2$ is in resonance with the scalaron oscillations (here, $a$ is the cosmological scale factor). Hence, for an increment in the comoving number density of photons in a small time interval $\Delta t$ during which the states in the momentum interval $\Delta p$ are filled, we have
\begin{equation}
\Delta n_\gamma = f (p) \Delta p = f (p) p \frac{\Delta a}{a} \, .
\end{equation}
Dividing this by $\Delta t$, we obtain the relation
\begin{equation}
\frac{d n_\gamma}{d t} = f (p) p H = 2 \Gamma n_\phi \, .
\end{equation}
Here, $\Gamma = \Gamma_{\phi\, \to\, \gamma\gamma}$, and $n_\phi$ is the comoving number density of scalarons. Hence,
\begin{equation} \label{fp}
f (p) = \frac{2 \Gamma n_\phi}{H_p p} \, , 
\end{equation}
where $H_p$ is the Hubble parameter at the time when electromagnetic modes with comoving momentum $p$ are in resonance, i.e., $p = a m/2$, which implies $1 + z = {1}/{a} = {m}/{2p}$  (we set $a_0 = 1$ today). In the late-time universe, we have, neglecting the energy density of radiation,
\begin{equation}
H = H_0 \sqrt{\Omega_\text{m} (1 + z)^3 + \Omega_\Lambda} \, .
\end{equation}
Using these relations in \eqref{fp}, we obtain
\begin{equation}\label{fp1}
f (p) = \frac{2 \Gamma n_\phi}{p H_0 \sqrt{\Omega_\text{m} \left( m / 2 p \right)^3 + \Omega_\Lambda}} = \frac{ q \overline \rho_\phi }{p H_0 \sqrt{\Omega_\text{m} \left( m / 2 p \right)^3 + \Omega_\Lambda}} \, ,
\end{equation}
where $\overline \rho_\phi = m n_\phi$ is the comoving energy density of the scalaron (equal to its current energy density), and [using estimate \eqref{Gamma1}]
\begin{equation}\label{q}
q = \frac{2 \Gamma}{m} \approx 0.7 \times 10^{- 50} \left( \frac{m}{\text{MeV}} \right)^2 \, .
\end{equation}

The total radiated photon number density is\footnote{Technically, we should integrate from $p = a_\text{rec} m / 2 = m / 2 (1 + z_\text{rec})$, where the subscript denotes the recombination epoch, beyond which the scattering and absorption of photons cannot be ignored. However, the integral converges rapidly in the infrared, so we can replace its lower limit by zero.}
\begin{equation} \label{ng}
n_\gamma = \frac{q \overline \rho_\phi}{H_0} \int_0^{m/2} \frac{d p}{p \sqrt{\Omega_\text{m} \left( m / 2 p \right)^3 + \Omega_\Lambda}} = \frac{q \overline \rho_\phi}{H_0} \times \frac{2 \arsinh \sqrt{1 / \Omega_\text{m} - 1}}{3 \sqrt{1 - \Omega_\text{m}}} \, ,
\end{equation}
where we have used the relation $\Omega_\Lambda = 1 - \Omega_\text{m}$. The last factor in the last expression in \eqref{ng} is approximately equal to 0.96 for the Planck-2018 value of $\Omega_\text{m} \approx 0.3$ \cite{Planck:2018vyg}, and we can replace it by unity. The spectral density \eqref{fp1} can then be written as
\begin{equation}
f (p) \, \approx \, \frac{n_\gamma}{p \sqrt{\Omega_\text{m} \left( m / 2 p \right)^3 + 1 - \Omega_\text{m}}} \, .
\end{equation}

We further have $\overline \rho_\phi = 2 M^2 H_0^2 \Omega_\phi$,  hence
\begin{equation}
n_\gamma \, \approx \, 2 q M^2 H_0 \Omega_\phi \, \approx \, 5.8 \times 10^{- 15} \left( \frac{m}{\text{MeV}} \right)^2 h_{70}^{-1} \, \text{cm}^{-3} \, ,
\end{equation}
where $h_{70} = H_0 / 70\, \text{km}\,\text{s}^{-1}\text{Mpc}^{-1}$, and we have used the Planck-2018 value for the dark-matter parameter $\Omega_\phi h_{70}^2 \approx 0.24$ \cite{Planck:2018vyg}.  

\begin{figure}
\begin{center}
\includegraphics[width=.7\textwidth]{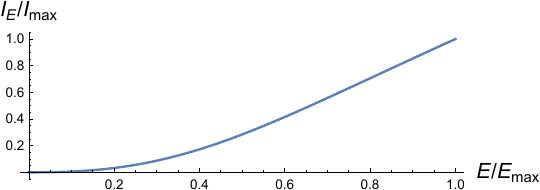}
\caption{Spectrum \eqref{IE} for $\Omega_\text{m} = 0.3$. \label{fig:spectrum}}
\end{center}
\end{figure}

This gives the estimate for the average intensity energy spectrum (see figure~\ref{fig:spectrum}):
\begin{equation} \label{IE}
I_E = \frac{E^2 f (E)}{4 \pi} = \frac{E n_\gamma}{4 \pi \sqrt{\Omega_\text{m} \left( m / 2 E \right)^3 + 1 - \Omega_\text{m}}} = \frac{I_\text{max}\, E / E_\text{max}}{\sqrt{\Omega_\text{m} \left( E_\text{max}/ E \right)^3 + 1 - \Omega_\text{m}}} \, ,
\end{equation}
where $E_\text{max} = m/2$ is the maximal photon energy, and 
\begin{equation}
I_\text{max} = \frac{m n_\gamma}{8 \pi} \simeq 8.5 \times 10^{-6} \left( \frac{m}{\text{MeV}} \right)^3 \frac{\text{MeV}}{\text{cm}^2~\text{s}~\text{sr}}
\end{equation} 
is the spectrum at this energy. 

As discussed previously in \cite{Cembranos:2008gj, Cembranos:2015svp}, dark-matter halos of compact objects (galaxies and clusters) should produce line emission at photon energy $E_\gamma = m/2$, the intensity of which would be greater the closer the scalaron mass is to the upper bound in \eqref{mup}, in view of the mass-dependence of ratio \eqref{q}.

\section{Thermal component of the scalaron}
\label{sec:thermal}

\subsection{Cross-sections}

Possible initial conditions for the scalaron dark matter were discussed in detail in our papers \cite{Shtanov:2022xew, Shtanov:2024nmf}. Their relation with the inflationary and preheating stages are hard to envision, firstly, because this issue is strongly model-dependent and, secondly, because at such high energy scales, the effective action for the theory may be qualitatively modified. In any case, it is assumed that, at reasonably low temperatures in the early universe, the scalaron field is either misaligned from its equilibrium value (as in the original scenario suggested in \cite{Cembranos:2008gj}), or evolves following the equilibrium value until the electroweak crossover, where it gets excited by the non-adiabatic evolution of the trace of the stress-energy tensor of matter \cite{Shtanov:2024nmf}.  This second scenario works only for a specific value of the scalaron mass, of the order of several meV\@. Both scenarios involve the formation of a homogeneous classical configuration (condensate) of the scalaron in the early universe, which then becomes dark matter. 

\begin{figure}
	\centering
\begin{tikzpicture}[scale=1.7]
\begin{feynman}
\vertex (a) at (0,.85);
\vertex (b) at (0,.15);
\vertex (c) at (-1,1) {$\mathrm{f}$};
\vertex (d) at (-1,0) {$\bar{\mathrm{f}}$};
\vertex (e) at (1,1) {$\phi$};
\vertex (f) at (1,0) {$\phi$};
\diagram* {
	 (e) -- [scalar] (a),
	 (f) -- [scalar] (b),
	(c) -- [fermion] (a) -- [fermion] (b) -- [fermion] (d),
};
\end{feynman} 
\end{tikzpicture} \qquad \qquad \qquad
\begin{tikzpicture}[scale=1.7]
\begin{feynman}
\vertex (a) at (0,.85);
\vertex (b) at (0,.15);
\vertex (c) at (-1,1) {g};
\vertex (d) at (-1,0) {g};
\vertex (e) at (1,1) {$\phi$};
\vertex (f) at (1,0) {$\phi$};
\diagram* {
	 (e) -- [scalar] (a),
	 (f) -- [scalar] (b),
	(c) -- [boson] (a) -- [boson] (b) -- [boson] (d),
};
\end{feynman} 
\end{tikzpicture}
\caption{Feynman graphs for the amplitude of the scalaron production by fermions (left) and gauge bosons (right). \label{fig:prod}}
\end{figure}
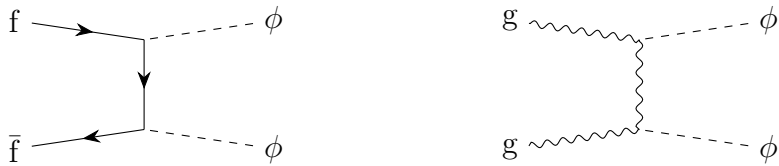

However, scalaron quanta can also be produced in the early hot universe through interactions with Standard Model particles. In this section, we estimate the contribution of this thermal component and demonstrate that it is negligible compared to the scalaron condensate. The processes for scalaron production are illustrated in figure~\ref{fig:prod}.

The cross-section of annihilation of fermions of mass $m_\text{f} \left( T \right)$ in thermal equilibrium into scalarons due to interactions \eqref{psicoup} is estimated as 
\begin{equation} \label{sigma-f}
\sigma_\text{f} \simeq 10^{-3} \left[ \frac{m_\text{f} \left( T \right)}{M} \right]^4 \frac{1}{T^2} \, ,
\end{equation}
where $T$ is the temperature of the cosmic plasma. For charged fermions, this expression holds for $T \gtrsim m_\text{f}\,$; below this temperature, fermions vanish from the plasma. For neutrinos with mass $m_\nu < m$, it is valid for $T \gtrsim m$, as neutrinos below this temperature lack sufficient energy to produce scalarons. 

The cross-section of annihilation of vector gauge bosons $W^\pm$ and $Z^0$ at temperatures below the electroweak crossover, where these fields are massive, from the direct interaction \eqref{gcoup} with the scalaron, is estimated to be
\begin{equation} \label{sigma-m}
\sigma_\text{b}^\text{dir} \simeq 10^{-3} \frac{T^2}{M^4} \, .
\end{equation}
A similar expression holds at all temperatures for the annihilation of Higgs bosons into two scalarons, arising from the interactions described by \eqref{Shn}.

The cross-section of annihilation of massless vector gauge bosons is estimated from the loop diagrams as
\begin{equation} \label{sigma-b}
\sigma_\text{b}^\text{an} \simeq 10^{-1} \left( \frac{\alpha_*}{M} \right)^4 T^2 \, .
\end{equation}
where $\alpha_* = g_*^2/ 4 \pi$, and $g_*$ is the gauge coupling constant. 

When the annihilating particles move with relativistic velocities, their relative velocity can be set to unity (the speed of light). The scalaron number density $n$, taking into account the expansion of the universe, then satisfies the Boltzmann equation
\begin{equation} \label{con-ev}
\frac{d \left( n a^3 \right)}{d t} = \sigma n_1 n_2 a^3 \, ,
\end{equation}
where $n_1$ and $n_2$ are the number densities of the colliding particles. It is convenient to proceed from the cosmological time to the cosmological temperature using the entropy conservation law $a^3 T^3 N = \text{const}$, where $N$ is the (temperature-dependent) effective number of degrees of freedom in thermal equilibrium. Using also the law
\begin{equation} \label{H}
H = \pi \sqrt{\frac{N}{60}} \frac{T^2}{M} \approx 0.4 \sqrt{N} \frac{T^2}{M} \, ,
\end{equation}
valid in a radiation-dominated universe, we can transform equation \eqref{con-ev} to
\begin{equation} \label{con-T}
\frac{d \left( n / \tau^3 \right)}{d \tau} = - \frac{\sqrt{N_i} T_i^2}{H_i} \frac{\sigma n_1 n_2 N^{1/6}}{\tau^6} = - \frac{\sqrt{N_f} T_f^2}{H_f} \frac{\sigma n_1 n_2 N^{1/6}}{\tau^6} \, ,
\end{equation}
where $\tau = N^{1/3} T$, and the indices ${i/f}$ denote the initial/final values of parameters. For the initial number density of the scalaron particles, we set $n_i = 0$.

\subsection{Annihilation of fermions}

In the case of annihilation of fermions, we have, counting the spin degrees of freedom of particles and antiparticles,
\begin{equation}
n_1 = n_2 = \frac{3 \zeta (3)}{2 \pi^2} T^3 \approx \frac16 T^3 \, .
\end{equation}
Substituting this into \eqref{con-T}, using \eqref{sigma-f}, and integrating this equation with the assumption of slow evolution of $N$, we obtain the scalaron number density $n_f$ at temperature $T_f$, where this process of production terminates:
\begin{equation}
n_f \approx 10^{-4} \left( \frac {m_\text{f}}{M} \right)^4 \frac{T_f^4}{H_f} \, , 
\end{equation}
where we have assumed $T_f \ll T_i$. Note that the integral is dominated by the lower integration limit $T = T_f$ in this case. The current number density of the scalarons produced by one fermion specie is then obtained by multiplying by a factor of $a_f^3 / a_0^3 \simeq 2 T_0^3 / N_f T_f^3$, where $T_0$ is the current cosmic microwave background temperature:
\begin{equation}
n_\text{(f)} \simeq 10^{-4} \left( \frac {m_\text{f}}{M} \right)^4 \frac{T_f T_0^3}{N_f H_f} \, .
\end{equation}

The final temperature $T_f$ will be roughly equal to the fermion mass (below this temperature, the fermion specie effectively vanish from thermal equilibrium). Using \eqref{H}, we obtain the final estimate
\begin{equation} \label{nof}
n_\text{(f)} \simeq \frac{10^{-4}}{N_f^{3/2}} \left( \frac {m_\text{f}}{M} \right)^3 T_0^3 \,.
\end{equation}
We observe that the heaviest fermion will have a major contribution to the scalaron number density. 

\subsection{Annihilation of bosons}

In the case of annihilation of bosons, we have estimates \eqref{sigma-m} and \eqref{sigma-b} for the cross section. As we will demonstrate shortly, the number density of thermal scalarons will be primarily determined by their production at high temperatures. We will use expression \eqref{sigma-b} for the cross-section, treating the vector bosons as massless.

We have, counting the spin degrees of freedom of vector bosons,
\begin{equation}
n_1 = n_2 = \frac{2 \zeta (3)}{\pi^2} T^3 \approx \frac14 T^3 \, .
\end{equation}
Substituting this into \eqref{con-T}, using \eqref{sigma-b}, and integrating this equation, we obtain the scalaron number density $n_f$ at temperature $T_f$, where the process of production terminates:
\begin{equation}
n_f \simeq 10^{-2} \left( \frac {\alpha_*}{M} \right)^4 \frac{N_f T_i^5 T_f^3}{N_i H_i} \, , 
\end{equation}
where we have also assumed $T_f \ll T_i$.  Note that the scalaron production is dominant at high temperatures $T_i$ in this case. The current number density of the scalarons produced by bosons is obtained by multiplying by a factor of $a_f^3 / a_0^3 \simeq 2 T_0^3 / N_f T_f^3$:
\begin{equation}\label{nob}
n_\text{(b)} \simeq 10^{-2} \left( \frac {\alpha_*}{M} \right)^4 \frac{T_i^5 T_0^3}{N_i H_i} \simeq 10^{-2} \frac{\alpha_*}{N_i^{3/2}} \left(\frac {\alpha_* T_i}{M} \right)^3 T_0^3 \, ,
\end{equation}
where we have used \eqref{H} in obtaining the last estimate.

The ratio of \eqref{nob} to \eqref{nof} gives
\begin{equation}\label{ratio}
\frac{n_\text{(b)}}{n_\text{(f)}} \simeq 10^2 \alpha_* \left( \frac{N_f}{N_i} \right)^{3/2} \left( \frac{\alpha_* T_i}{m_\text{f}} \right)^3 \, .
\end{equation}
The initial temperature $T_i$ will be roughly equal to the reheating temperature, which is typically very large, reaching values of order $10^{12}$--$10^{15}$~GeV, so that ratio \eqref{ratio} is typically much larger than unity. In this case, production of scalarons due to bosons will strongly dominate.

The estimates for the scalaron production due to Higgs boson annihilation will be the same with $\alpha_*$ replaced by $10^{-1/2}$. All our estimates are valid up to a numerical factor.

\subsection{Fraction of thermal scalarons}

Since thermal scalarons are primarily produced by bosons, we estimate their contribution to the total scalaron number density that constitutes dark matter. The current total number density of dark matter scalarons is calculated as
\begin{equation} \label{nc}
n_\text{s} = \frac{\rho_\text{\tiny DM}}{m} \, ,
\end{equation}
where $\rho_\text{\tiny DM}$ is the cosmological energy density of dark matter, and $m$ is the scalaron mass. Taking the ratio of \eqref{nob} to \eqref{nc}, and using the fact that $T_0^4 \approx 1.5 \rho_\gamma$, we obtain
\begin{equation}
\frac{n_\text{(b)}}{n_\text{s}} \simeq \frac{m}{\rho_\text{\tiny DM}} \times 10^{-2} \frac{\alpha_*}{N_i^{3/2}} \left(\frac {\alpha_* T_i}{M} \right)^3 T_0^3 \approx  10^{-2} \frac{\alpha_*}{N_i^{3/2}} \frac{m}{T_0} \left(\frac {\alpha_* T_i}{M} \right)^3 \frac{\rho_\gamma}{\rho_\text{\tiny DM}} \, .
\end{equation}
Using the observational estimate $\rho_\gamma / \rho_\text{\tiny DM} \approx 2 \times 10^{-4}$, we obtain
\begin{equation}
\frac{n_\text{(b)}}{n_\text{s}} \simeq 10^{-6} \frac{\alpha_*}{N_i^{3/2}} \frac{m}{T_0} \left(\frac {\alpha_* T_i}{M} \right)^3 \, .
\end{equation}

For the numerical estimates, we take into account that the highest reasonable reheating temperature is $T_i \sim 10^{15}~\text{GeV} \sim 10^{-3} M$, $T_0 \approx 2 \times 10^{-4}~\text{eV}$, $\alpha_* \sim 10^{-1}$, $N_i \simeq 100$, and the highest allowable scalaron mass is $m \simeq 1~\text{MeV}$. Hence, we obtain the upper estimate
\begin{equation}
\frac{n_\text{(b)}}{n_\text{s}} \lesssim 10^{- 12} \, .
\end{equation}
The contribution from Higgs boson annihilation is estimated to be at most two orders of magnitude higher. However, this remains a negligible fraction, indicating that scalaron radiation from the hot plasma in the early universe can be safely ignored.

\section{Discussion}
\label{sec:summary}

It is remarkable that a light scalaron in generic $f(R)$ gravity can serve as a candidate for dark matter \cite{Cembranos:2008gj, Cembranos:2015svp, Shtanov:2021uif, Shtanov:2022xew, Shtanov:2024nmf}. An advantage of this theory can be seen in its economy: it introduces no new fundamental fields, aside from the fact that the metric acquires an additional degree of freedom. When assuming minimal coupling of the Standard Model to the metric field, the theory has a single essential free parameter\,---\,the mass of the scalaron\,---\,which makes it highly predictive. However, a comprehensive description of the scalaron’s interaction with matter in the Standard Model has been lacking in the literature, prompting us to fill this gap.

In this paper, we revisited the effective one-loop coupling between the scalaron and the electromagnetic field within the Standard Model, providing a detailed derivation of the scalaron decay rate into two photons [Eq.~\eqref{Gamma}], confirming the result of \cite{Cembranos:2008gj, Cembranos:2015svp}. We then used this result to compute the average background radiation generated by these decays. This decay rate can also be used to search for the potential line emission from dark matter halos predicted by this model \cite{Cembranos:2008gj, Cembranos:2015svp}. Our analysis addresses the existing controversy in the literature concerning this interaction, which was also described in the initial version of this paper \cite{Shtanov:2025nue-1}. We have shown that directly calculating loop diagrams with proper regularisation eliminates the ambiguities present in approaches that rely on Jacobians from field redefinitions.

The dark-matter model under consideration assumes the formation of a homogeneous classical condensate of the scalaron in the early universe, which subsequently evolves into dark matter. However, the early hot universe can radiate thermal scalaron quanta due to its interaction with Standard Model particles. We have estimated the contribution of this thermal component to the total energy density of the scalaron and shown that it is negligible compared to the energy density of the classical condensate.

A key challenge in this theory is its connection to the inflationary epoch and preheating, during which the initial conditions for the scalaron condensate must be set. This issue is highly model-dependent and warrants further investigation, as the effective action may undergo qualitative changes at such high energy scales. We intend to address these topics in greater detail in future work.

\section*{Acknowledgements}

We are grateful to Eduard Gorbar and Valery Gusynin for valuable discussions. The work of Y.~Shtanov was funded by the National Academy of Sciences of Ukraine under project 0121U109612, by the National Research Foundation of Ukraine under project 2023.03/0149, and by the Simons Foundation.

\appendix

\section{Effective interaction with photons}
\label{app:decay}

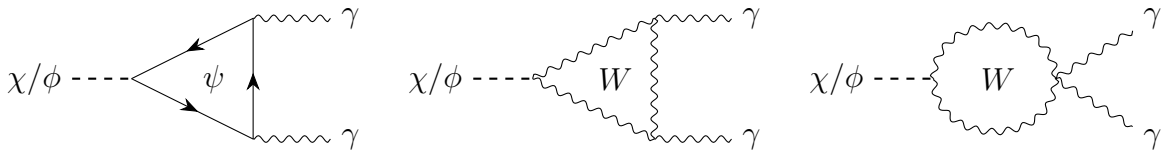
\begin{figure}
\centering
\begin{tikzpicture}[scale=1.6]
\begin{feynman}
\vertex (a) at (.2,0) {$\chi/\phi$};
\vertex (b) at (1,0);
\vertex (c) at (2,-.5);
\vertex (d) at (2,.5);
\vertex (e) at (2.8,-.5) {$\gamma$};
\vertex (f) at (2.8,.5) {$\gamma$};
\vertex (g) [label=0:$\psi$] at (1.5,0);
\diagram* { (a) -- [thick, scalar] (b), 
(b) -- [fermion] (c) -- [fermion] (d) -- [fermion] (b), (c) -- [photon] (e), (d) -- [photon] (f)};
\end{feynman}
\end{tikzpicture} \hspace{3pt} 
\begin{tikzpicture}[scale=1.6]
\begin{feynman}
\vertex (a) at (.2,0) {$\chi/\phi$};
\vertex (b) at (1,0);
\vertex (c) at (2,-.5);
\vertex (d) at (2,.5);
\vertex (e) at (2.8,-.5) {$\gamma$};
\vertex (f) at (2.8,.5) {$\gamma$};
\vertex (g) [label=0:$W$] at (1.45,0);
\diagram* { (a) -- [thick, scalar] (b), 
(b) -- [boson] (c) -- [boson] (d) -- [boson] (b), (c) -- [photon] (e), (d) -- [photon] (f)};
\end{feynman}
\end{tikzpicture} \hspace{3pt} 
\begin{tikzpicture}[scale=1.6]
\begin{feynman}
\vertex (a) at (.2,0) {$\chi/\phi$};
\vertex (b) at (1,0);
\vertex (c) at (2,0);
\vertex (d) at (2.8,-.5) {$\gamma$};
\vertex (e) at (2.8,.5) {$\gamma$};
\vertex (f) [label=0:$W$] at (1.3,0);
\diagram* { (a) -- [thick, scalar] (b), 
(b) -- [boson, half right] (c) -- [boson, half right] (b), (d) -- [photon] (c) -- [photon] (e)};
\end{feynman}
\end{tikzpicture}
\caption{Feynman diagrams in unitary gauge contributing to the decay of the Higgs boson $\chi$ or the scalaron $\phi$ into two photons $\gamma$.  \label{fig:decay}}	
\end{figure}

We recall that the effective one-loop interaction of the Higgs boson $\chi$ with photons in the Standard Model has the form
\begin{equation}
{\cal L}_{\text{\tiny H}\gamma \gamma} = - \frac{\alpha }{8 \pi} F \left( m_\text{\tiny H} \right) \frac{\chi}{v} F_{\mu\nu} F^{\mu\nu} \, . 
\end{equation}
The factor $F (m)$ for arbitrary mass $m$ is given by \cite{Shifman:1979eb, Marciano:2011gm} 
\begin{equation}\label{F}
F (m) = F_W \left( \beta_W \right) + \sum_\psi  Q_\psi^2 F_f \left( \beta^{}_\psi \right) \, .
\end{equation}
Here, the sum is taken over all fermionic fields $\psi$ with the account of the colour of quarks, $Q_\psi$ is the electric charge in units of the positron's charge, $\beta_i = 4 m_i^2 / m^2$, and
\begin{align}
F_W (\beta) &= 2 + 3 \beta + 3 \beta (2 - \beta) f^2 (\beta) \, , \\[2pt]
F_f (\beta) &= - 2 \beta \left[ 1 + (1 - \beta) f^2 (\beta) \right] \, , \label{Ff}
\end{align}
\begin{equation}\label{f}
f (\beta) = \left\{ 
\begin{array}{cl} 
\arcsin \beta^{-1/2}  &\ \ \text{for} \ \beta \geq 1 \, , \\[3pt]
\dfrac12 \left( \pi + {\rm i} \ln \dfrac{1 + \sqrt{1 - \beta}}{1 - \sqrt{1 - \beta}} \right)  &\ \ \text{for} \ \beta < 1 \, .
\end{array} 
\right.
\end{equation}
This effective interaction is responsible for the Higgs-boson decay into two photons, and the relevant Feynman diagrams for this process are shown in figure~\ref{fig:decay}.

We now use reasoning similar to that in \cite{Goldberger:2007zk}. As demonstrated in section~\ref{sec:coup}, the scalaron interactions \eqref{psicoup} and \eqref{gcoup} to first order in $\phi$ are identical to those of the Higgs boson, with the substitution $\chi / v \to - \phi / 2 M$. Since the Higgs boson enters the diagrams of figure~\ref{fig:decay} only at a single vertex, we have, from similar diagrams for the scalaron,
\begin{equation}\label{fff}
{\cal L}_{\phi \gamma \gamma} = \frac{\alpha}{16 \pi} F (m)  \frac{\phi}{M} F_{\mu\nu} F^{\mu\nu} \, , 
\end{equation}
where $F$, in which the argument now is the scalaron mass $m$, is given by \eqref{F}--\eqref{f}. The function $F (m)$ is real for $m \leq 2 m_e$. We have $F \left( 2 m_e \right) \approx - 4.2$, and $F (m) \to - 11/3$ as $m \to 0$. In the formal limit of all $m_i \to 0$ (at the electroweak crossover), only the contribution from the $W^\pm$ bosons survive, and we have $F \to 2$.

The total width of the scalaron decay into two photons is equal to
\begin{equation}\label{Gamma}
\Gamma_{\phi\, \to\, \gamma\gamma} = \frac{\alpha^2 m^3}{2^{10} \pi^3 M^2} \left| F \right|^2 \, .
\end{equation}
In the neighbourhood of $m = 2 m_e$, equation \eqref{Gamma} gives
\begin{equation}
\Gamma_{\phi\, \to\, \gamma\gamma} \approx 5.2 \times 10^{- 30} \left( \frac{m}{\text{MeV}} \right)^3\, \text{s}^{-1} \approx \left[ 1.9 \times 10^{29} \left( \frac{\text{MeV}}{m} \right)^3\, \text{s}\, \right]^{-1} \, .
\end{equation}

\begin{figure}
\begin{center}
\includegraphics[width=.75\textwidth]{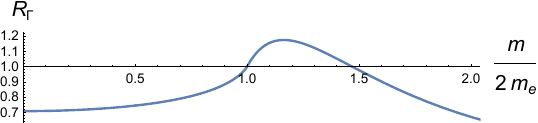}
\caption{Ratio \eqref{rat} as a function of the scalaron mass $m$. \label{fig:Gamma}}
\end{center}
\end{figure}

This equation is valid with a good precision over the entire possible range of the scalaron mass.  In figure~\ref{fig:Gamma}, we show the dimensionless ratio
\begin{equation} \label{rat}
R_\Gamma = \frac{\Gamma ( m ) / m^3}{\Gamma ( 2 m_e ) / ( 2 m_e )^3} 
\end{equation}
as a function of $m$, where $\Gamma ( m )$ is the exact decay width $\Gamma_{\phi\, \to\, \gamma\gamma}$ as a function of $m$. It is observed that this ratio is close to unity over a wide range of scalaron masses.

\section{Field redefinition and renormalisation}
\label{app:redef}

Here we demonstrate that we can consistently apply the Dirac-field redefinition \eqref{psicon} when computing the fermionic loop diagrams shown in figure~\ref{fig:decay}. In terms of the original electrically charged Dirac field, the corresponding Lagrangian in the Einstein frame is
\begin{align} \label{split}
{\cal L}_\psi &= \frac12 e^{- {3 \phi}/{2 M}} \left[ \overline \psi\! \stackrel{\rightarrow}{\widetilde D}\! \psi - \overline \psi\! \stackrel{\leftarrow}{\widetilde D}{\!\!}^\dagger \psi \right] - m_\psi e^{- {2 \phi}/{M}}\, \overline \psi \psi \nonumber \\[3pt] &= \frac12 e^{- {3 \phi}/{2 M}} \left[ \overline \psi  \bigl( \stackrel{\rightarrow}{\widetilde D} - m_\psi \bigr) \psi - \overline \psi \bigl( \stackrel{\leftarrow}{\widetilde D}{\!\!}^\dagger + m_\psi \bigr) \psi \right] - m_\psi \left( e^{- {2 \phi}/{M}} - e^{- {3 \phi}/{2 M}} \right) \overline \psi \psi \, , 
\end{align}
where $\widetilde D = {\rm i} \gamma^a \widetilde e_a^\mu \widetilde D_\mu$ is the covariant Dirac operator in the Einstein metric frame involving the electromagnetic field $A_\mu$, the arrow indicates the direction of its action, and $\widetilde e_a^\mu$ is the tetrad field of the Einstein metric frame. We argue that the first term (enclosed in square brackets) in the final expression of Eq.~\eqref{split} does not contribute to the loop diagram and can therefore be subjected to an arbitrary multiplicative redefinition of the spinor field. This implies that the theory satisfies the field-redefinition equivalence theorems \cite{Chisholm:1961tha, Kamefuchi:1961sb, Divakaran:1963yxz, Arzt:1993gz, Cohen:2024fak}, at least at the one-loop level, leaving the $S$-matrix invariant\,---\,even though the interaction involves derivatives of the spinor.

The first term in \eqref{split} is proportional to the field equation for the free Dirac spinor; hence, it replaces one of the propagators in the triangular diagram by a constant. As a result, each of the triangular diagrams produced by this term, though quadratically divergent in four dimensions, depends only on one of the two photon momenta, $k_1$ or $k_2$. Hence, it is proportional to the scalar product $\varepsilon_1 \cdot \varepsilon_2$ of the photon polarisation vectors. On the other hand, in dimensional regularisation, which respects the gauge invariance, it should be proportional to $\left( k_1 \cdot k_2 \right) \left( \varepsilon_1 \cdot \varepsilon_2 \right) - \left( \varepsilon_1 \cdot k_2 \right) \left( \varepsilon_2 \cdot k_1 \right)$. This is possible only if such a diagram vanishes identically.

This can be verified by direct computation: all diagrams arising from the terms in square brackets on the second line of Eq.~\eqref{split}, after a shift of the loop momentum integration variable, yield the same expression, proportional to
\begin{equation} \label{dim}
\int d^n k \left[ \frac{2 \left( \varepsilon_1 \cdot k \right) \left( \varepsilon_2 \cdot k \right)}{\left( k^2 - m_\psi^2 + {\rm i} \epsilon \right)^2} - \frac{\varepsilon_1 \cdot \varepsilon_2}{k^2 - m_\psi^2 + {\rm i} \epsilon} \right] \, .
\end{equation}
The two integrals in this expression cancel each other in dimensional regularisation.  Pauli--Villars regularisation has a similar effect. The subtracted terms introduced by this method have the same form \eqref{dim} as the original diagram but differ in sign and in the fermion mass. The shift in the integration momentum is applied uniformly across all terms and is independent of the fermion masses. The sum of these terms yields an integrand whose integral is now absolutely convergent. This combined integral can then be evaluated using dimensional regularisation, where each individual term contributes zero\,---\,resulting in a vanishing total.

Only the final (mass) interaction term in Eq.~\eqref{split} remains, and it is not affected by field redefinitions to linear order in $\phi$. This term is responsible for the contribution \eqref{Ff} to the effective interaction \eqref{fff}. This direct diagrammatic calculation using regularisation, without the need to introduce counterterms, has the advantage of being unambiguous and independent of the choice of conformal frame.

In the previous preprint version of this paper \cite{Shtanov:2025nue-1}, we adopted the approach of \cite{Katsuragawa:2016yir}, which involves calculation of the regularised functional Jacobian for fermions following the field redefinition in \eqref{psicon}. This resulted in an anomalous contribution of the form
\begin{equation}\label{an}
{\cal L}_{\phi \gamma \gamma}^\text{an} = \frac{\alpha}{16 \pi} F_\text{an} \frac{\phi}{M} F_{\mu\nu} F^{\mu\nu} 
\end{equation}
to the effective interaction, where $F_\text{an}$ is a certain constant. Such a contribution apparently reflects the freedom to add an arbitrary finite counterterm of this form to the effective action during renormalisation, and this freedom is fixed by the prescription of Fujikawa's regularisation of the Jacobian. It now appears that this calculation is subject to several issues.

To begin with, the constant $F_\text{an}$ in \eqref{an} depends on the choice of the original spinor field variables in the functional integration measure. To illustrate this, we again consider the case of electrodynamics in the Jordan frame, with the action
\begin{equation} \label{spinor}
S \left[ e_a^\mu, \psi \right] = \int \left[ \frac{1}{2} \left( \overline \psi\! \stackrel{\rightarrow}{D}\! \psi - \overline \psi\! \stackrel{\leftarrow}{D}{\!\!}^\dagger \psi \right) - m_\psi \overline \psi \psi \right] \sqrt{-g}\, d^4 x \, , 
\end{equation}
where $D$ and $e_a^\mu$ are, respectively, the Dirac operator and tetrad field of the Jordan metric frame. As is well known, the spinor field variables in a diffeomorphism-invariant functional integration measure are not the original $\psi$ but the spinor semidensities $\psi' = (- g)^{1/4} \psi$ (see \cite{Fujikawa:2004cx}). So let us write the action in terms of these variables: 
\begin{equation}\label{spinor'}
S' \left[ e_a^\mu, \psi^{\prime\,} \right] = \int \left[ \frac{1}{2} \left( \overline \psi{}'\! \stackrel{\rightarrow}{D}\! \psi' - \overline \psi{}'\! \stackrel{\leftarrow}{D}{\!\!}^\dagger \psi' \right) - m_\psi \overline \psi{}' \psi' \right] d^4 x \, .
\end{equation}
Note that the flat-space limits of actions \eqref{spinor} and \eqref{spinor'} are identical.

After the transformation $e_a^\mu = e^{\phi / 2 M}\, \widetilde e_a^\mu$ to the Jordan frame, action \eqref{spinor'} becomes
\begin{equation}\label{spinorE'}
S' \left[ e^{\phi / 2 M}\, \widetilde e_a^\mu, \psi^{\prime\,} \right] = \int \left[ \frac{1}{2} e^{\phi / 2 M} \left( \overline \psi{}'\! \stackrel{\rightarrow}{\widetilde D}\! \psi' - \overline \psi{}'\! \stackrel{\leftarrow}{\widetilde D}{\!\!}^\dagger \psi' \right) - m_\psi \overline \psi{}' \psi' \right] d^4 x \, .
\end{equation}
Comparing the Lagrangian in this expression with the first line of \eqref{split}, we see that different spinor field redefinitions are required in the two cases to eliminate the scalaron from the spinor kinetic term. For \eqref{split}, the appropriate redefinition is given by \eqref{psicon}, whereas for \eqref{spinorE'}, the redefinition $\psi' = e^{- \phi / 4 M} \widetilde \psi{}'$ must be used. After these transformations, the two Lagrangians become equivalent in the flat-space limit of the Einstein frame. However, the transformations yield different formal Jacobians in the functional integral. When regularised in Euclidean space using Fujikawa's method \cite{Fujikawa:2004cx}, which preserves gauge invariance, the logarithms of the corresponding Jacobians in the flat-space limit take the following forms\footnote{The fermions in the Standard Model belong to chiral representations of the electroweak gauge group. In this connection, there is ongoing controversy in the literature regarding the presence of parity-violating terms of the form ${\cal F}^{\alpha\beta} {\cal F}^{\mu\nu} \epsilon_{\alpha \beta \mu \nu}$ in the chiral determinant (see \cite{Larue:2023tmu, Larue:2023qxw, Larue:2024zen, Bonora:2024imk}). In any case, such terms would cancel for interactions of the scalaron with photons and gluons, whose couplings to fermions are chirality-symmetric.}:
\begin{align} \label{J}
\ln J &= \ln \left| \Det \frac{{\cal D} \psi}{{\cal D} \widetilde \psi} \right|^{-2} = - \frac32 \Tr \frac{\phi}{M} =  - \frac{Q^2}{16 \pi^2} \frac{\phi}{M} F_{\mu\nu} F^{\mu\nu} \, , \\
\ln J' &= \ln \left| \Det \frac{{\cal D} \psi'}{{\cal D} \widetilde \psi{}'} \right|^{-2} = \frac12 \Tr \frac{\phi}{M} =  \frac{Q^2}{48 \pi^2} \frac{\phi}{M} F_{\mu\nu} F^{\mu\nu} \, , \label{canon}
\end{align}
where $Q$ is the fermion electric charge. Therefore, all else being equal, the resulting anomalous interactions of the scalaron with the electromagnetic field will differ between these two frameworks.\footnote{Incidentally, it is the result in equation \eqref{canon} that correctly captures the coupling of the scalaron to the anomalous trace of the stress-energy tensor (see \cite{Fujikawa:2004cx}).}

We believe that the resolution of this puzzle lies in the proper regularisation of quantum electrodynamics, and we present here a heuristic argument in support of this view. One of the most transparent ways to implement Pauli--Villars regularisation with a single subtraction is to augment the Lagrangian by introducing two ghost fields: a bosonic spinor field $\zeta$, and a massive bosonic vector ghost field $G_\mu$, with corresponding field strength $G_{\mu\nu}$ \cite{Schwartz:2014sze}. The vector ghost couples to both fermions in the same way as the gauge field $A_\mu$, but has negative Lagrangian. The resulting extended action takes the form
\begin{align} \label{spinor-full}
&S \left[ e_a^\mu, \ldots \right] = \int \left[ \frac{1}{2} \left( \overline \psi\! \stackrel{\rightarrow}{D}\! \psi - \overline \psi\! \stackrel{\leftarrow}{D}{\!\!}^\dagger \psi \right) - m_\psi \overline \psi \psi - \frac14 F_{\mu\nu} F^{\mu\nu} \right] \sqrt{-g}\, d^4 x \nonumber \\ &\qquad + \int \left[ \frac{1}{2} \left( \overline \zeta\! \stackrel{\rightarrow}{D}\! \zeta - \overline \zeta\! \stackrel{\leftarrow}{D}{\!\!}^\dagger \zeta \right) - M_\zeta \overline \zeta \zeta + \frac14 G_{\mu\nu} G^{\mu\nu} - \frac12 M_G^2\, G_\mu G^\mu \right] \sqrt{-g}\, d^4 x \, . 
\end{align}

Now we proceed to the Einstein metric frame in this action by making the conformal transformation $e_a^\mu = e^{\phi / 2 M}\, \widetilde e_a^\mu$, where $\widetilde e_a^\mu$ is the tetrad field of the Einstein frame. Apart from the mass term of the vector ghost, the vector field part is conformally invariant, and we omit it. Action \eqref{spinor-full} for the spinor fields then becomes
\begin{align} \label{spinorE}
S \left[ e^{\phi / 2 M}\, \widetilde e_a^\mu, \psi, \zeta \right] = \int \left[ \frac{1}{2} e^{- 3 \phi / 2 M} \left( \overline \psi\! \stackrel{\rightarrow}{\widetilde D}\! \psi - \overline \psi\! \stackrel{\leftarrow}{\widetilde D}{\!\!}^\dagger \psi \right) - e^{- 2 \phi / M} m_\psi \overline \psi \psi \right] \sqrt{- \widetilde g}\, d^4 x \nonumber \\ + \int \left[ \frac{1}{2} e^{- 3 \phi / 2 M} \left( \overline \zeta\! \stackrel{\rightarrow}{\widetilde D}\! \zeta - \overline \zeta\! \stackrel{\leftarrow}{\widetilde D}{\!\!}^\dagger \zeta \right) - e^{- 2 \phi / M} M_\zeta \overline \zeta \zeta \right] \sqrt{- \widetilde g}\, d^4 x \, . 
\end{align}
Here, the tildes denote metric-related variables in the Einstein frame.

The scalaron field can be eliminated from the kinetic terms of the spinors by the spinor field redefinitions
\begin{equation}
\psi = e^{3 \phi / 4 M} \widetilde\psi \, , \qquad \zeta = e^{3 \phi / 4 M} \widetilde\zeta \, .
\end{equation}
Each of these redefinitions yields a Jacobian in the quantum functional integral. However, the spinors involved have opposite statistics. As a result, the corresponding Jacobians are mutual inverses (at least at one loop), so their product yields a total Jacobian equal to unity. After the field redefinition, the action for the spinors becomes
\begin{align} \label{spinorEt}
\widetilde S \left[ \phi, \widetilde e_a^\mu, \widetilde \psi, \widetilde \zeta \right] = \int \left[ \frac{1}{2} \left( \overline {\widetilde \psi}\! \stackrel{\rightarrow}{\widetilde D}\! \widetilde \psi - \overline {\widetilde \psi}\! \stackrel{\leftarrow}{\widetilde D}{\!\!}^\dagger \widetilde \psi \right) - e^{- \phi / 2 M} m_\psi \overline {\widetilde \psi} \widetilde \psi \right] \sqrt{- \widetilde g}\, d^4 x \nonumber \\ + \int \left[ \frac{1}{2} \left( \overline {\widetilde \zeta}\! \stackrel{\rightarrow}{\widetilde D}\! \widetilde \zeta - \overline {\widetilde \zeta}\! \stackrel{\leftarrow}{\widetilde D}{\!\!}^\dagger \widetilde \zeta \right) - e^{- \phi / 2 M} M_\zeta \overline {\widetilde \zeta}\, \widetilde \zeta \right] \sqrt{- \widetilde g}\, d^4 x \, . 
\end{align}
Note that the field redefinition has been performed consistently, the scalaron is no longer coupled to the kinetic terms, and the Pauli--Villars ghost now correctly regularises the ultraviolet divergences in loop diagrams in the Einstein frame. 

This line of reasoning breaks down when more than one Pauli--Villars subtraction is required to regularise the loop integrals, as in such cases the subtractions cannot be implemented at the Lagrangian level. Nevertheless, it offers additional support for the conclusion we previously reached using dimensional regularisation.

\bibliographystyle{JHEP}
\bibliography{scalaron}

\end{document}